\documentclass[conference]{IEEEtran}
\IEEEoverridecommandlockouts
\usepackage{cite}
\usepackage{amsmath,amssymb,amsfonts}
\usepackage{algorithmic}
\usepackage{graphicx}
\usepackage{textcomp}
\usepackage{xcolor}
\usepackage{subcaption}
\usepackage{csquotes}
\usepackage{url}
\usepackage[linesnumbered,ruled,vlined]{algorithm2e}

\def\BibTeX{{\rm B\kern-.05em{\sc i\kern-.025em b}\kern-.08em
    T\kern-.1667em\lower.7ex\hbox{E}\kern-.125emX}}
\begin{document}

\title{CyberDep: Towards the Analysis of Cyber-Physical Power System Interdependencies Using Bayesian Networks and Temporal Data\\

\thanks{This work was supported by the Sandia Laboratory Directed Research and Development Project \#229324 and the US Department of Energy under award DE-CR0000018.
}
}

\author{
  Leen Al Homoud\IEEEauthorrefmark{1}\IEEEauthorrefmark{4}, \emph{Student Member, IEEE},
  Katherine Davis\IEEEauthorrefmark{1}, \emph{Senior Member, IEEE},\\
  Shamina Hossain-McKenzie\IEEEauthorrefmark{3}, \emph{Member, IEEE},
  Nicholas Jacobs\IEEEauthorrefmark{3}, \emph{Member, IEEE}, \\
  \IEEEauthorblockA{
    \IEEEauthorrefmark{1}Department of Electrical and Computer Engineering, Texas A\&M University, College Station, TX, USA \\
    \IEEEauthorrefmark{3}Sandia National Laboratories, Albuquerque, NM, USA
  }
  \IEEEauthorrefmark{4}leen.alhomoud@ieee.org
  }

\IEEEoverridecommandlockouts
\IEEEpubid{\makebox[\columnwidth]{979-8-3503-7240-3/24/\$31.00 \copyright2024 IEEE\hfill} \hspace{\columnsep}\makebox[\columnwidth]{ }}

\maketitle

\begin{abstract}
Modern-day power systems have become increasingly cyber-physical due to the ongoing developments to the grid that include the rise of distributed energy generation and the increase of the deployment of many cyber devices for monitoring and control, such as the Supervisory Control and Data Acquisition (SCADA) system. Such capabilities have made the power system more vulnerable to cyber-attacks that can harm the physical components of the system. As such, it is of utmost importance to study both the physical and cyber components together, focusing on characterizing and quantifying the interdependency between these components. This paper focuses on developing an algorithm, named \emph{CyberDep}, for Bayesian network generation through conditional probability calculations of cyber traffic flows between system nodes. Additionally, \emph{CyberDep} is implemented on the temporal data of the cyber-physical emulation of the WSCC 9-bus power system. The results of this work provide a visual representation of the probabilistic relationships within the cyber and physical components of the system, aiding in cyber-physical interdependency quantification.

\end{abstract}


\section{Introduction}

Over the past decade, power systems have become increasingly recognized as cyber-physical systems. The importance of understanding the interdependency between cyber and physical components is highlighted by the many new developments in the grid, such as, but not limited to, renewable energy integration, distributed energy generation monitoring and control, and new cyber-security technology. Specifically, a power grid's Supervisory Control and Data Acquisition (SCADA) system allows for the monitoring and control of physical components in the system and communicates with devices through a variety of protocols, such as the Distributed Network Protocol 3 (DNP3) \cite{dnp3}. This protocol is one of the most widely used in electric power utilities. Such developments have made the power grid more vulnerable to cyber-attacks that target the physical components of the system at the generation, transmission, and distribution levels. Two infamous threats are the Industroyer and Industroyer2 malware that affected Ukraine in 2016 and 2022, respectively \cite{ukraine}. Both of these threats targeted electrical substations in the country, with the Industroyer malware sending SCADA commands to the field devices resulting in an hour-long power outage across Ukraine.

As such, it is now of utmost importance to analyze and study the power system as both a cyber and physical system, while taking into account how the cyber and physical components of the system are interdependent. There is a lot of emerging literature focused on understanding cyber-physical power system interdependencies \cite{chen2020robustness, marashi2021identification, huang2014small, marashi2017consideration}. In \cite{chen2020robustness}, the authors study the interdependency relationship between the physical power grid and its corresponding communication network when dealing with and mitigating cascading failures. Through numerical simulations of a cyber-attack on a cyber-physical power system model, the authors found that power systems divide into clusters when facing cascading failures. These results showed that there is a correlation between system robustness and cluster size, proving that these cyber-physical clusters are still interdependent of each other, but operating separately. In \cite{marashi2021identification}, the authors claim that \enquote{interdependence is an intrinsic feature of cyber-physical systems.} The authors back up this claim by characterizing cyber-physical interdependencies using correlation metrics aimed at predicting the propagation of failure following a cyber-attack on the network. Huang et al. \cite{huang2014small} also study interdependencies concerning cascading failures following a mathematical estimation approach using concepts of graph theory. Other applications that utilize cyber-physical interdependency analysis in power grids include improving power system reliability modeling \cite{marashi2017consideration} and developing cyber-physical resiliency metrics \cite{venkataramanan2020cp,clark2017cyber}.

In addition to the literature detailed above, Bayesian Networks have also been used in cyber-physical power systems for many applications including, but not limited to, attack graph generation \cite{sahu2021structural}, cyber threat mitigation \cite{zebrowski2022bayesian}, scalable anomaly detection \cite{krishnamurthy2014scalable}, and risk analysis and assessment \cite{almajali2020risk, lyu2020bayesian}. Sahu et al. \cite{sahu2021structural} focused on developing a Bayesian attack graph and updating it through the use of constraint-based structural learning methods that focus on scalability and accuracy. In \cite{zebrowski2022bayesian}, the authors develop a quantitative framework using Bayesian networks to define all possible vulnerabilities and optimize this framework to achieve mitigation of cyber-physical attacks, while Krishnamurthy et al. \cite{krishnamurthy2014scalable} focused on creating Bayesian networks of power systems to study the different cyber-physical relations between the nodes to achieve anomaly detection focused on power system scalability.

It is also important to note that this work is part of a larger effort aimed at characterizing and quantifying cyber-physical power system interdependencies \cite{hossain2023towards, jacobs2024leveraging, sun2023bio}. Much of the ongoing work is focused on the development and use of a variety of graph clustering methods that aid in characterizing cyber and physical disturbances and cyber-physical interdependencies. Therefore, the work in this paper focuses on continuing these efforts through the development of a Bayesian network generation algorithm that inputs temporal data generated from the earlier work in \cite{hossain2022harmonized} and outputs graph visualizations of the probabilistic relationship between different nodes in a cyber-physical power system model. 

With that being said, the contributions of this paper are as follows:

\begin{enumerate}
    \item Development of a Bayesian network generation algorithm through the use of temporal data and conditional probability calculations of cyber traffic flows between system nodes.
    \item Application of this algorithm on the temporal data of the cyber-physical emulation of the WSCC 9-bus system \cite{wscc} under physical, cyber, and cyber-physical disturbances.
    \item Visualization of the probabilistic relationships between the different system nodes, aiding in cyber-physical interdependency quantification.
\end{enumerate}

\section{Methodology}

In this section, we first describe the temporal dataset generated by the earlier work in \cite{hossain2022harmonized} and list the physical, cyber, and cyber-physical threat vectors that make up the different disturbance scenarios. Then, we focus on conceptualizing Bayesian networks and detailing the development of the generation of such networks, specifically known as Dependency Graphs. 

\subsection{Disturbances and Dataset Description}

The HARMONIE \cite{hossain2022harmonized} project focused on developing a cyber-physical response engine that generates real-time cyber-physical power system mitigations through a machine learning classification framework and automated remedial action schemes (RAS). The techniques developed were tested in a cyber-physical emulation environment built using a real-time digital simulator (RTDS) and SCEPTRE\texttrademark \cite{sceptre}. SCEPTRE\texttrademark \cite{sceptre} is a modeling and emulation platform developed by Sandia for emulating Industrial Control Systems (ICS). It allows for the modeling and emulation of different virtual and hardware devices, such as, but not limited to, switches, servers, and relays. It also supports power system simulations and ICS communication protocols such as DNP3.

\begin{figure}
    \centering
    \includegraphics[width=1\linewidth]{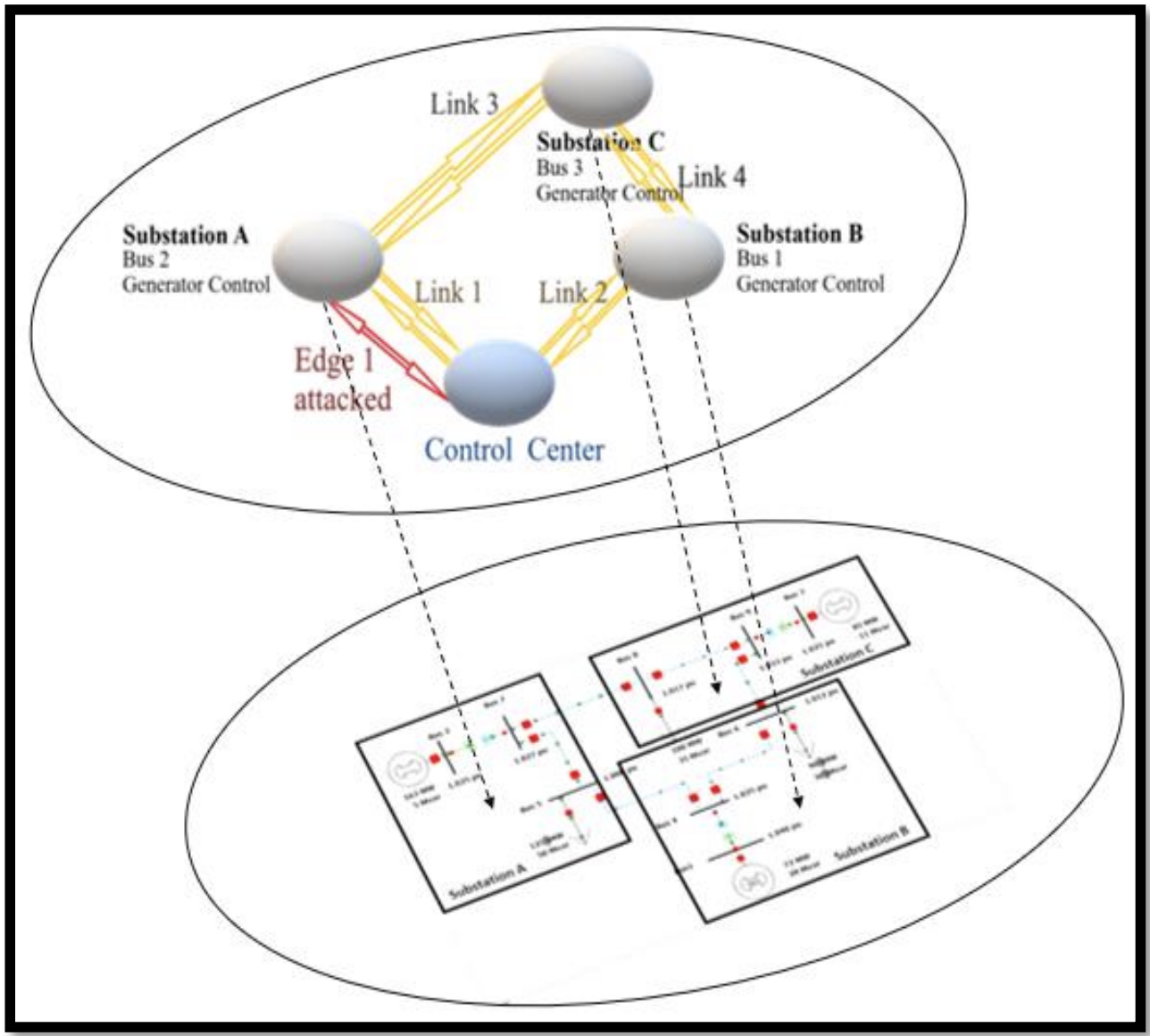}
    \caption{Cyber-physical mapping of the WSCC 9-bus system \cite{hossain2022harmonized}.}
    \label{fig:cp_mapping}
\end{figure}

\begin{figure}
    \centering
    \includegraphics[width=1\linewidth]{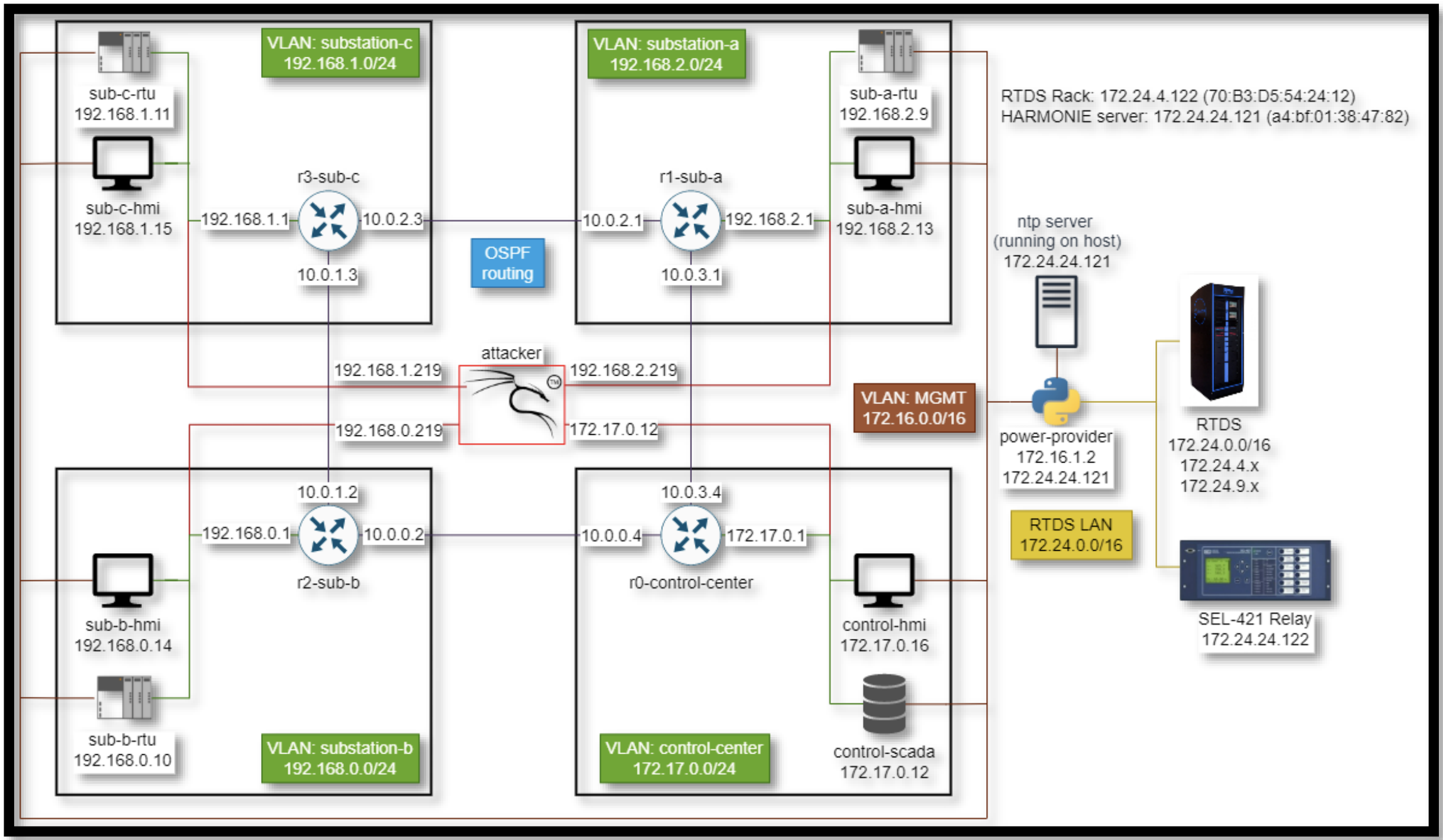} 
    \caption{Network diagram of the WSCC 9-bus emulation environment \cite{hossain2022harmonized}.}
    \label{fig:network}
\end{figure}

The data used in this paper was generated as part of the different experiments that were run in the emulation environment on the WSCC 9-bus power system \cite{wscc}, where the cyber-physical mapping of the 9-bus system is shown in Figure \ref{fig:cp_mapping} and the network diagram is shown in Figure \ref{fig:network}. In Figure \ref{fig:cp_mapping}, it can be observed that the WSCC 9-bus system is divided into three substations, with Substation A containing Bus 2, Substation B containing Bus 1 (slack bus), and Substation C containing Bus 3. The fourth substation in the cyber-physical emulation is the control center, which contains the SCADA system that sends commands to the field devices. In the environment, the WSCC system is emulated as a 4-substation network, with a router connecting each substation to the rest of the network, as shown in Figure \ref{fig:network}.

The three disturbances that were tested in this environment are \cite{hossain2022harmonized}:

\begin{enumerate}
    \item Cyber: The cyber disturbance consisted of a Denial-of-Service (DOS) intrusion. The mitigation implemented for this threat included using firewall rules to block adversary communication.
    \item Physical: The physical disturbance included the loss of a generator and a branch that led to line overloading. The mitigation implemented is load shedding at two different buses using an automated remedial action schemes (AutoRAS) algorithm \cite{hossain2022harmonized,li2020investigation,li2021towards}. 
    \item Cyber-Physical: This disturbance is a combination of the above disturbances with both mitigation strategies implemented.
\end{enumerate}

The dataset contains four cyber-physical disruption scenarios based on the three disturbances listed above. These scenarios are run three times each and are as follows:

\begin{enumerate}
    \item Baseline: Normal system operations.
    \item DOS: This scenario includes only the cyber disturbance with no physical disruptions.
    \item No Mitigation: This scenario includes a physical disturbance as well as a cyber one that affects load shedding.
    \item Mitigation: This is the same as the No Mitigation scenario with an addition of the firewall rules put in place to block the cyber-attack.
\end{enumerate}

\subsection{Dependency Graph Generation}

Dependency graphs (DGs) are a type of Bayesian network that helps represent the different cyber and physical system characteristics during normal operating conditions and under threats. Dependency graph (DG) conceptualization is provided in \cite{vellaithurai2014cpindex}, where the authors focused on developing a cyber-physical resiliency metric using graph theory concepts. For this paper, we will focus on DGs to help quantify cyber-physical interdependencies. DGs are generated through the conditional probability calculations of the frequency of communication between the different nodes using the following formula \cite{vellaithurai2014cpindex}:

\[
P(x|P(x)) = 1 - \prod_{p^i_x \in P(x)} \left(1 - \mathbf{1}_{(p_x^i)} \times P(p_x^i \rightarrow x)\right)
\]

where,
\begin{align*}
    P(x|P(x))&:\text{Probability of \emph{x} given \emph{P(x)}} \\
    \mathbf{1}_{(p_x^i)}&:\text{Indicator function, which is 1 if the condition} \\
    & \text{in parentheses holds and 0 otherwise} \\
    P(p_x^i \rightarrow x)&:\text{Probability of information flow from $p_x^i$ to \emph{x}}
\end{align*}



A DG captures the relationships between the different files and processes in a network, which depend on whether there is data flow between two nodes. As such, a DG implies that if there is traffic moving from object $o_i$ to $o_j$, then object $o_j$ is dependent on object $o_i$. This dependency is represented by an edge on the graph, $o_i \rightarrow o_j$. In this example, the dependency relationship is characterized by three components, which are the source, $o_i$, the sink, $o_j$, and the security contexts, cyber traffic information between nodes $o_i$ and $o_j$. The nodes of a DG are modeled as binary random variables, and the edges are labeled with the frequency of communication between two different nodes, which is the calculated probability dependent on the number of system calls between each of the nodes.

\begin{figure}
    \centering
    \includegraphics[width=0.85\linewidth]{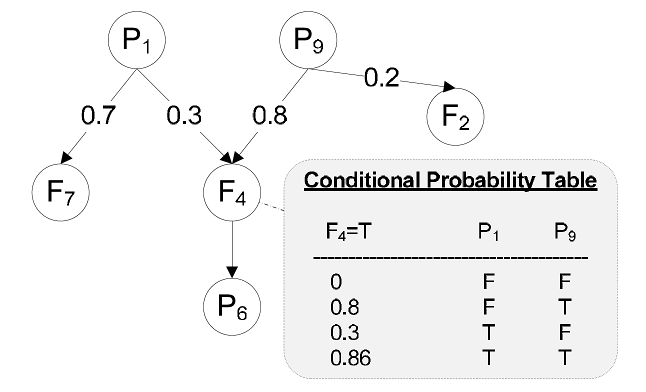}
    \caption{Sample dependency graph obtained from \cite{vellaithurai2014cpindex}.}
    \label{fig:DGexample}
\end{figure}

System calls, $syscalls$ in short, are the communication requests and responses made between each node. For the DNP3 communication protocol, the $syscalls$ under consideration are Request Link Status, Read, Respond, and Direct Operate commands, explained in more detail in the DNP3 protocol primer \cite{dnp3}. A sample of a dependency graph can be seen in Figure \ref{fig:DGexample} \cite{vellaithurai2014cpindex}. The conditional probability that File F4 would be affected if a cyber-attack were to affect either Process P1 or P9 is given by \cite{vellaithurai2014cpindex}: 

\[
P(F4|P1, P9) = 1 - (1 - 0.3) \times (1 - 0.8) = 0.86
\]

Similarly, the probability that File F2 would be affected if Process P9 is affected is 0.2, and the probability that File F7 would be affected if Process P1 is affected is 0.7. 

Algorithm \ref{algo:example} shows the steps for generating the DGs. The datasets collected from the emulation are in JSON file format. The first step is to load the files, and then filter the traffic out for DNP3 data. DNP3 traffic is selected because it collects information on the physical components of the network as well as cyber information, thus providing a better insight into cyber-physical interdependencies. Once the input data is processed, the IP addresses are then mapped to the device names using the network topology information. The frequency of communication is then counted, and the conditional probability is then calculated. Four graphs for each of the runs are generated, as a result, for each of the four scenarios. 

\begin{algorithm}
    \caption{Dependency Graph Generation}
    \label{algo:example}
    
    \KwData{Raw temporal data collected in \cite{hossain2022harmonized}.}
    \KwResult{Four graphs per experiment run that visualize the probabilistic relationship between the different nodes in the system.}

    Load JSON files of temporal data.\
    
    Filter for DNP3 traffic data.\
    
    Convert JSON files to CSV format.\
    
    Map IP addresses with device names using the network topology.\
    
    Count the number of times each component communicated with the control center's SCADA system.\
    
    Calculate the conditional probability to quantify the frequency of communication between the nodes.\
    
    Plot the graphs for each of the four scenarios for all three runs and label the edges with the probabilities.\
\end{algorithm}

\section{Results and Discussions} 

\begin{figure*}
    \centering

    \begin{subfigure}[b]{0.44\textwidth}
        \includegraphics[width=\textwidth]{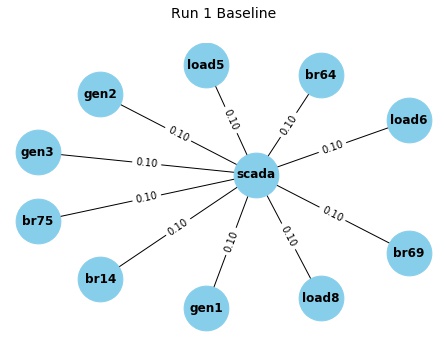}
        \caption{Baseline results for Run 1.}
        \label{fig:sub1.1}
    \end{subfigure}
    \hfill
    \begin{subfigure}[b]{0.44\textwidth}
        \includegraphics[width=\textwidth]{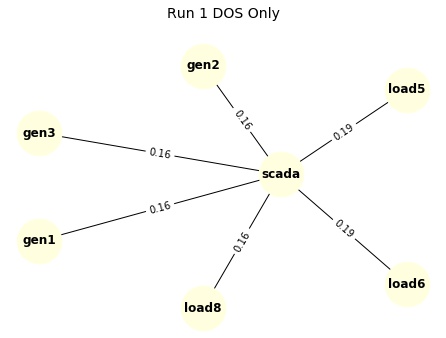}
        \caption{DOS results for Run 1.}
        \label{fig:sub2.1}
    \end{subfigure}

    \vspace{\baselineskip}

    \begin{subfigure}[b]{0.44\textwidth}
        \includegraphics[width=\textwidth]{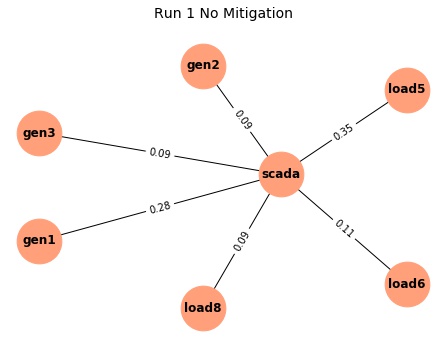}
        \caption{No mitigation results for Run 1.}
        \label{fig:sub3.1}
    \end{subfigure}
    \hfill
    \begin{subfigure}[b]{0.44\textwidth}
        \includegraphics[width=\textwidth]{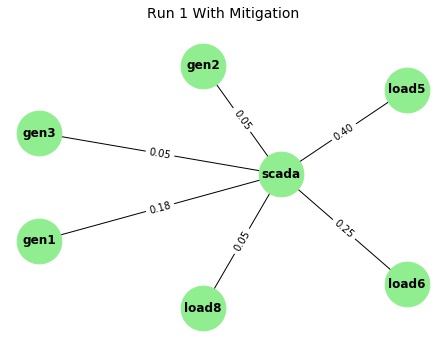}
        \caption{Mitigation results for Run 1.}
        \label{fig:sub4.1}
    \end{subfigure}

    \caption{Results for all four scenarios for the first run of the experiments in the dataset. The edges are labeled with the probabilities calculated using the temporal data.}
    \label{fig:main.1}
\end{figure*}

\begin{figure}
    \centering

    \begin{subfigure}[b]{0.37\textwidth}
        \includegraphics[width=\textwidth]{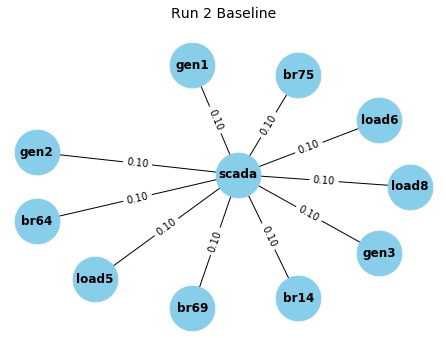}
        \caption{Baseline results for Run 2.}
        \label{fig:sub1.2}
    \end{subfigure}
    \hfill
    \begin{subfigure}[b]{0.37\textwidth}
        \includegraphics[width=\textwidth]{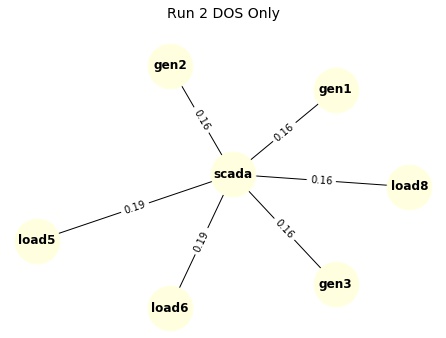}
        \caption{DOS results for Run 2.}
        \label{fig:sub2.2}
    \end{subfigure}

    \vspace{\baselineskip}

    \begin{subfigure}[b]{0.37\textwidth}
        \includegraphics[width=\textwidth]{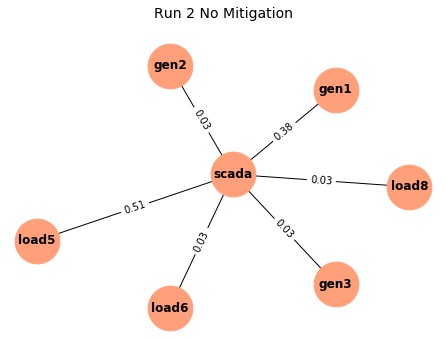}
        \caption{No mitigation results for Run 2.}
        \label{fig:sub3.2}
    \end{subfigure}
    \hfill
    \begin{subfigure}[b]{0.37\textwidth}
        \includegraphics[width=\textwidth]{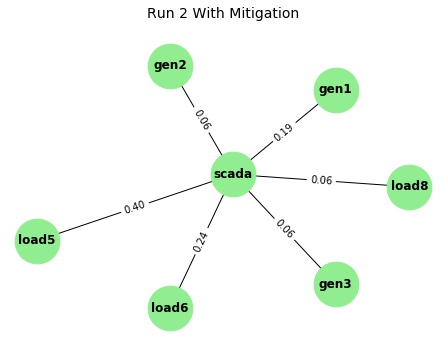}
        \caption{Mitigation results for Run 2.}
        \label{fig:sub4.2}
    \end{subfigure}

    \caption{Results for all four scenarios for the second run of the experiments in the dataset. The edges are labeled with the probabilities calculated using the temporal data.}
    \label{fig:main.2}
\end{figure}

\begin{figure}
    \centering

    \begin{subfigure}[b]{0.37\textwidth}
        \includegraphics[width=\textwidth]{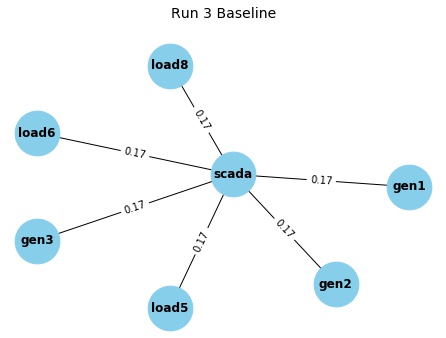}
        \caption{Baseline results for Run 3.}
        \label{fig:sub1.3}
    \end{subfigure}
    \hfill
    \begin{subfigure}[b]{0.37\textwidth}
        \includegraphics[width=\textwidth]{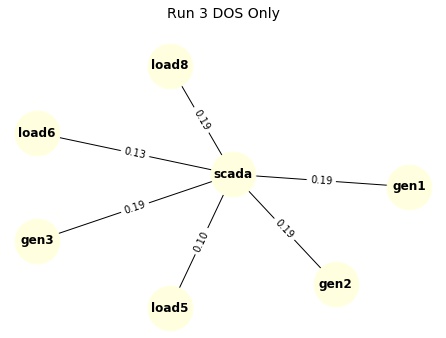}
        \caption{DOS results for Run 3.}
        \label{fig:sub2.3}
    \end{subfigure}

    \vspace{\baselineskip}
    
    \begin{subfigure}[b]{0.37\textwidth}
        \includegraphics[width=\textwidth]{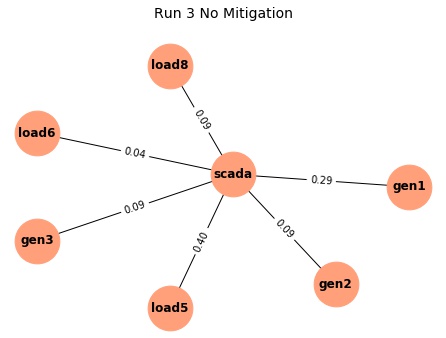}
        \caption{No mitigation results for Run 3.}
        \label{fig:sub3.3}
    \end{subfigure}
    \hfill
    \begin{subfigure}[b]{0.37\textwidth}
        \includegraphics[width=\textwidth]{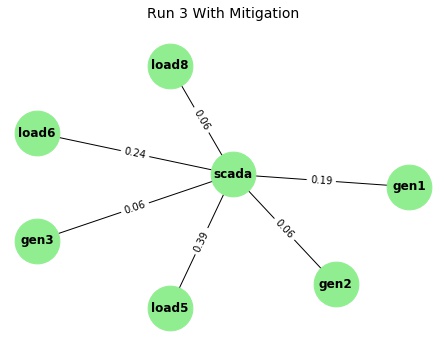}
        \caption{Mitigation results for Run 3.}
        \label{fig:sub4.3}
    \end{subfigure}

    \caption{Results for all four scenarios for the third run of the experiments in the dataset. The edges are labeled with the probabilities calculated using the temporal data.}
    \label{fig:main.3}
\end{figure}

In this section, we will discuss the results for each of the three experimental runs and their respective four cyber-physical disruption scenarios. Figures \ref{fig:main.1}, \ref{fig:main.2}, and \ref{fig:main.3} display the results for runs 1, 2, and 3, respectively.


Across all three experimental runs, the baseline graphs exhibited similar patterns of behavior in which the probabilities of all the edges were equal. The probabilities amounted to 0.1 for each edge in run 1 and run 2 and 0.17 for each edge in run 3. A crucial observation is that the DOS Only, No Mitigation, and With Mitigation scenarios behaved similarly across runs 1 and 2; however, the DOS Only scenario in run 3 behaved differently. For runs 1 and 2, the DOS Only scenario shows that the highest probabilities are for the edges connecting each of loads 5 and 6 to the SCADA node. This result makes sense as this scenario consists of a DOS threat through DNP3 increasing the amount of packets traveling between the objects affected. It is also important to note that while this intrusion was cyber in nature, we were able to see the relation and the effect of a purely cyber-attack on two physical components in the network, loads 5 and 6. For the DOS Only scenario for run 3, the opposite is observed. The edges connecting loads 5 and 6 to the SCADA node have the lowest probabilities. This could be due to the fact that the DOS threat in this scenario was not implemented for the same duration of time as the other two runs. As such, further analysis of the network topology and experimental setup would be required to interpret this result. 


Moving onto the No Mitigation and With Mitigation scenarios for all three runs, similar patterns and probabilities were observed for the edges in the graphs. Specifically, an important observation can be seen where the edges with the highest probabilities are the ones connecting generator 1 (the slack bus) and load 5 to the SCADA node. These results are also justified as this scenario consists of both a cyber-attack and a physical disturbance to the system affecting load shedding. What can be understood from this graph is that the DOS cyber-attack is implemented on load 5 in the network, and the physical disruption affected the load shedding setup in the power system that the SCADA node needed to send more commands to change the generation values at the slack bus (generator 1) to make up for the loss or increase in power generation. These are all valid points to discuss as, once again, the cyber-physical interdependencies can be understood from the dependency graphs (DGs). 

Last but not least, observing the With Mitigation results for all runs shows us that the edges with the highest probabilities are the ones connecting loads 5 and 6 to the SCADA node with generator 1 having the second highest probability on the edge connecting it to the SCADA node. These results also make sense as there are firewall rules set up now that prevent the cyber-attack from occurring, hence the increased communication between the SCADA node and the rest of the objects to prevent the attack from occurring. 

\section{Conclusions and Future Work}

In conclusion, \emph{CyberDep} was developed to generate dependency graphs using the temporal data of the WSCC 9-bus system and perform conditional probability calculations of cyber traffic flows between system nodes. Additionally, we can observe from the results above that the work on \emph{CyberDep} aided in providing insight into cyber-physical interdependencies through quantifying and visualizing the probabilistic relationships between the different system nodes.

Future work includes the consideration of bi-directional traffic flows, integration of more datasets to include additional cyber and physical devices, such as routers and switches, and expansion to larger power systems. \emph{CyberDep} can be utilized to infer and build access paths for cyber-physical threat models and generate cyber-physical kill chains.


\section{Acknowledgments}
The authors would like to thank Christopher Goes at Sandia National Laboratories for his efforts in generating the datasets used in this work and the members of Sandia Laboratory Directed Research and Development Project \#229324 for their collaborative discussions. This work was supported by the Sandia Laboratory Directed Research and Development Project \#229324 and the US Department of Energy under award DE-CR0000018.


\bibliographystyle{IEEEtran}
\bibliography{ref}

\vfill

\end{document}